\documentclass{llncs}
\usepackage{cite}
\usepackage[utf8]{inputenc}

\usepackage{graphicx}
\usepackage{color}
\usepackage{latexsym}
\usepackage{amsfonts}
\usepackage{crayola}
\usepackage{hyperref}

\title{Towards a format for describing networks}
\titlerunning{Towards a format for describing networks}
\subtitle{2. Format elements}
\author{Vladimir Batagelj\inst{1,2,3,5}\orcidID{0000-0002-0240-9446} 
   \and\protect\\ Tomaž Pisanski\inst{1,2}\orcidID{0000-0002-1257-5376}
   \and Iztok Savnik\inst{1}\orcidID{0000-0002-3994-4805}
   \and\protect\\ Ana Slavec\inst{1,4}\orcidID{0000-0002-0171-2144}
   \and Nino Bašić\inst{1,2}\orcidID{0000-0002-6555-8668}
}
\authorrunning{abbreviated author list}
\institute{UP FAMNIT Koper
   \and IMFM Ljubljana
   \and UL FMF Ljubljana
   \and InnoRenew CoE Izola
   \and \email{vladimir.batagelj@fmf.uni-lj.si}
   \and \url{https://github.com/bavla/netsJSON}
}

\newcommand{\clock}{\count254=\time \divide\count254 by 60
 \count255=\count254 \multiply\count255 by -60
 \advance\count255 by \time
 \ifnum\count254<10 0\fi\number\count254\,:\,%
 \ifnum\count255<10 0\fi\number\count255}

\newcommand{\keyw}[1]{\textcolor{red}{\emph{#1}}}

\newcommand{\network}[1]{\mathcal{#1}}
\newcommand{\vertices}[1]{\mathcal{#1}}
\newcommand{\edges}[1]{\mathcal{#1}}

\newcommand{\functions}[1]{\mathcal{#1}}

\newcommand{\Mw}{\mathop{\raisebox{-1.5pt}{\mbox{$\Box$\kern-.55em\raisebox{2.5pt}{{\tiny $r$}}\kern2.9pt}}}}
\newcommand{\Mv}{\mathop{\raisebox{-1.5pt}{\mbox{$\Box$\kern-.55em\raisebox{2.5pt}{{\tiny $h$}}\kern2.9pt}}}}

\graphicspath{{./},{../pics/}}
\begin{document}
\maketitle
\begin{abstract}
The key elements that a common format for describing networks should include are discussed.

\end{abstract}

\section{Introduction}
In 2023, the International Network for Social Network Analysis (INSNA) requested that Zachary Neal form a working group to develop \textbf{recommendations for sharing network data and materials}. They were published in \textit{Network Science} in 2024 \cite{Neal_et_al._2024} accompanied with the
\textit{Endorsement page} \cite{Endorsement}.

It would be useful to have a common “archiving/intermediate” format that can describe (almost) all networks and supports authoring, deposit, exchange, visualization, reuse, and preservation of network data.  It is easy to write converters from this format to a selected format or corresponding network reading procedures.

\subsection{Software support for network analysis}


There are many tools and programs for network analysis UCINET, Pajek, Gephi, NetMiner, Cytoscape, NodeXL, E-Net, Tulip, PUCK, GraphViz, SocNetV, Kumu, Polinode, etc.

Programmers can use network analysis packages/libraries in different programming languages 
\begin{itemize}
\item \textbf{Python:} NetworkX, igraph, Snap.py, graph-tool, NetworKit, PyGraphistry, Nets, cdlib, node2vec, DGL, PyG, Tulip, PyVis, 
\item \textbf{R:} igraph, statnet, sna, qgraph, RSiena, tnet, multiplex, NetSim, influenceR, tidygraph, intergraph, netUtils, ggraph, networkD3, visNetwork, DiagrammeR, graphlayouts, ndtv, 
\item \textbf{Julia:} LightGraphs, Graphs, MetaGraphs, SimpleWeightedGraphs, Erdos, MultilayerGraphs, GraphDataFrameBridge, GraphIO, NetworkDynamics, TemporalGPs, EcologicalNetwork, CommunityDetection, GraphPlot, NetworkLayout, 
\item \textbf{C++:} Boost Graph Library, igraph, SNAP, NetworKit, NetworkX, Graph-tool, GraphX, GraphBLAS, Lemon Graph Library, GraphHopper, Gelly, Tulip, OGDF,
\item etc.
\end{itemize}

They are supporting different network description formats: CSV, UCINET DL, Pajek NET, Gephi GEXF, GDF, GML, GraphML, GraphX, GraphViz DOT, Tulip TPL, Netdraw VNA, Spreadsheet, etc. \cite{GephiFormats}.

In addition, network data appears in several application areas such as chemistry and genealogy. There are many formats for describing molecular graphs: Molfile, SDF, CML, PDB, XYZ, CIF, FASTA, CDX, CDXML, JCAMP-DX, SMILES, InChI, and others. The most widely used format for genealogical data exchange, GEDCOM is a plain text file format that stores information about individuals, families, events, and sources. It has several derivatives. It is considered an exchange format between various genealogy programs, which are often based on their own format. Some of the most well-known are Ancestry Tree Files, Family Tree Maker, Legacy Family Tree, RootsMagic, OpenGen Alliance, Open Archives Format, FamilySearch JSON, Gramps XML, TEI, PROGEN, Webtrees, PAF.



\subsection{Network representations}

There are three commonly used file representations of graphs and networks.
\begin{itemize}
\item \textbf{Link list (with weights)}
This is the most commonly used and expressively most flexible representation.
\item \textbf{Matrix representation}
It is often found in older sources. It is suitable for describing smaller, denser simple networks. We lose the distinction between directed and undirected and multiple links. For larger networks, which are usually sparse, it requires a lot of space -- most matrix entries have the value 0.
\item \textbf{Neighbor sets}
This representation is very economical but only useful for networks without link properties.
\end{itemize}

\subsection{Repositories of graph and network data}

Network datasets are available in multiple repositories
\begin{itemize} 
\item \href{https://icon.colorado.edu/#!/networks}{ICON} -- Colorado Index of Complex Networks
\item \href{https://sites.google.com/site/ucinetsoftware/datasets}{UCINET} datasets
\item \href{https://github.com/bavla/Nets/tree/master/data}{Pajek} networks
\item \href{https://networkdata.ics.uci.edu/index.html}{UCI Network Data Repository}
\item \href{http://www.casos.cs.cmu.edu/tools/data2.php}{CASOS} -- \textbf{C}omputational \textbf{A}nalysis of \textbf{S}ocial and \textbf{O}rganizational \textbf{S}ystems
\item \href{https://snap.stanford.edu/data/}{SNAP} -- \textbf{S}tanford \textbf{N}etwork \textbf{A}nalysis \textbf{P}latform
\item \href{http://konect.cc/}{KONECT} -- \textbf{Ko}blenz \textbf{Ne}twork \textbf{C}ollec\textbf{t}ion
\item \href{https://networks.skewed.de/}{Netzschleuder} -- network catalogue, repository and centrifuge
\item \href{https://github.com/schochastics/networkdata}{Schochastics} network data
\item \href{https://networkrepository.com/}{Network Repository}
\item \href{https://www.stats.ox.ac.uk/~snijders/siena/siena_datasets.htm}{Siena data sets}
\item \href{https://houseofgraphs.org/}{The House of Graphs}
\item \href{http://atlas.gregas.eu/}{Encyclopedia of Graphs}
\item \href{https://graphsym.net/}{Datasets of Highly Symmetric Objects}
\item \href{https://github.com/briatte/awesome-network-analysis?tab=readme-ov-file#datasets}{Awesome Network Analysis}  datasets
\item \href{https://kintip.net/que-faisons-nous/kinsources}{Kinsources}
\item \href{https://www.rcsb.org/}{RCSB Protein Data Bank }
\end{itemize}
\noindent
None of these repositories explicitly adhere to FAIR data principles.

Some repositories only offer metadata about an individual network and a link to the actual dataset, while others also store the data itself and offer the user a display of basic network characteristics.

\section{Description of traditional networks}

\subsection{Description of networks using a spreadsheet}

How to describe a network $\network{N}=(\vertices{V},\edges{L},\functions{P},\functions{W})$?
In principle the answer is simple -- we list its components $\vertices{V}$, $\edges{L}$, $\functions{P}$, and $\functions{W}$.
The simplest way is to describe a network $\network{N}$ by providing $(\vertices{V},\functions{P})$ and 
$(\edges{L},\functions{W})$ in a form of two tables.

As an example, let us describe a part of the network determined by the bibliographical data about the following works:
\href{http://www.cambridge.org/tw/academic/subjects/sociology/sociology-general-interest/generalized-blockmodeling}{Generalized blockmodeling}, \href{http://link.springer.com/article/10.1007%2FBF02293706}{Clustering with relational constraint}, \href{http://www.sciencedirect.com/science/article/pii/S0378873308000397}{Partitioning signed social networks}, \href{http://www.journals.uchicago.edu/doi/abs/10.1086/225469}{The Strength of Weak Ties} 
already used in the first part of this paper.

There are nodes of different types (modes): persons, papers, books, series, journals, publishers; and different relations among them:  author\_of, editor\_of, contained\_in, cites, published\_by. For some types of nodes additional properties are known: sex, year, volume, number, first and last page, etc.

Both tables are often maintained in Excel. They can be exported as text in \href{https://en.wikipedia.org/wiki/Comma-separated_values}{CSV} (Comma Separated Values) format. Tables for our example are given in Figures 1 and 2. In large networks, we split a network into some subnetworks -- a collection, to avoid the empty cells.


\begin{figure}
{\renewcommand{\baselinestretch}{0.7}\scriptsize
\begin{verbatim}
name;mode;country;sex;year;vol;num;fPage;lPage;x;y
"Batagelj, Vladimir";person;SI;m;;;;;;809.1;653.7
"Doreian, Patrick";person;US;m;;;;;;358.5;679.1
"Ferligoj, Anuška";person;SI;f;;;;;;619.5;680.7
"Granovetter, Mark";person;US;m;;;;;;145.6;660.5
"Moustaki, Irini";person;UK;f;;;;;;783.0;228.0
"Mrvar, Andrej";person;SI;m;;;;;;478.0;630.1
"Clustering with relational constraint";paper;;;1982;47;;413;426;684.1;380.1
"The Strength of Weak Ties";paper;;;1973;78;6;1360;1380;111.3;329.4
"Partitioning signed social networks";paper;;;2009;31;1;1;11;408.0;337.8
"Generalized Blockmodeling";book;;;2005;24;;1;385;533.0;445.9
"Psychometrika";journal;;;;;;;;741.8;086.1
"Social Networks";journal;;;;;;;;321.4;236.5
"The American Journal of Sociology";journal;;;;;;;;111.3;168.9
"Structural Analysis in the Social Sciences";series;;;;;;;;310.4;082.8
"Cambridge University Press";publisher;UK;;;;;;;534.3;238.2
"Springer";publisher;US;;;;;;;884.6;174.0
\end{verbatim}
}
\caption{File \textbf{\texttt{bibNodes.csv}} -- $(\vertices{V},\functions{P})$ table for nodes}
\end{figure}


\begin{figure}
{\renewcommand{\baselinestretch}{0.7}
\scriptsize
\begin{verbatim}
from;relation;to
"Batagelj, Vladimir";authorOf;"Generalized Blockmodeling"
"Doreian, Patrick";authorOf;"Generalized Blockmodeling"
"Ferligoj, Anuška";authorOf;"Generalized Blockmodeling"
"Batagelj, Vladimir";authorOf;"Clustering with relational constraint"
"Ferligoj, Anuška";authorOf;"Clustering with relational constraint"
"Granovetter, Mark";authorOf;"The Strength of Weak Ties"
"Granovetter, Mark";editorOf;"Structural Analysis in the Social Sciences"
"Doreian, Patrick";authorOf;"Partitioning signed social networks"
"Mrvar, Andrej";authorOf;"Partitioning signed social networks"
"Moustaki, Irini";editorOf;"Psychometrika"
"Doreian, Patrick";editorOf;"Social Networks"
"Generalized Blockmodeling";containedIn;"Structural Analysis in the Social Sciences"
"Clustering with relational constraint";containedIn;"Psychometrika"
"The Strength of Weak Ties";containedIn;"The American Journal of Sociology"
"Partitioning signed social networks";containedIn;"Social Networks"
"Partitioning signed social networks";cites;"Generalized Blockmodeling"
"Generalized Blockmodeling";cites;"Clustering with relational constraint"
"Structural Analysis in the Social Sciences";publishedBy;"Cambridge University Press"
"Psychometrika";publishedBy;"Springer"
\end{verbatim}
}
\caption{File \textbf{\texttt{bibLinks.csv}} -- $(\vertices{L},\functions{W})$ table for links}
\end{figure}



\begin{figure}
{\renewcommand{\baselinestretch}{0.7}
\scriptsize
\begin{verbatim}
# transforming CSV file to Pajek files, by Vladimir Batagelj, June 2016
colC <- c(rep("character",4),rep("numeric",5)); nas=c("","NA","NaN")
nodes <- read.csv2("bibNodes.csv",encoding='UTF-8',colClasses=colC,na.strings=nas)
n <- nrow(nodes); M <- factor(nodes$mode); S <- factor(nodes$sex)
mod <- levels(M); sx <- levels(S); S <- as.numeric(S); S[is.na(S)] <- 0
links <- read.csv2("bibLinks.csv",encoding='UTF-8',colClasses="character")
F <- factor(links$from,levels=nodes$name,ordered=TRUE)
T <- factor(links$to,levels=nodes$name,ordered=TRUE)
R <- factor(links$relation); rel <- levels(R)
net <- file("bib.net","w"); cat('*vertices ',n,'\n',file=net)
clu <- file("bibMode.clu","w"); sex <- file("bibSex.clu","w")
cat('%',file=clu); cat('%',file=sex)
for(i in 1:length(mod)) cat(' ',i,mod[i],file=clu)
cat('\n*vertices ',n,'\n',file=clu)
for(i in 1:length(sx)) cat(' ',i,sx[i],file=sex)
cat('\n*vertices ',n,'\n',file=sex)
for(v in 1:n) {
  cat(v,' "',nodes$name[v],'"\n',sep='',file=net);
  cat(M[v],'\n',file=clu); cat(S[v],'\n',file=sex)
}
for(r in 1:length(rel)) cat('*arcs :',r,' "',rel[r],'"\n',sep='',file=net)
cat('*arcs\n',file=net)
for(a in 1:nrow(links))
  cat(R[a],': ',F[a],' ',T[a],' 1 l "',rel[R[a]],'"\n',sep='',file=net)
close(net); close(clu); close(sex)
\end{verbatim}
}
\caption{\textbf{\texttt{CSV2Pajek.R}} -- program for converting tables into network in Pajek format }
\end{figure}


\subsection{Factorization and description of large networks}

To save space and improve computing efficiency we often replace values of categorical variables with integers. In R this encoding is called a \keyw{factorization}.

We enumerate all possible values of a given categorical variable (coding table) and afterward replace each value with the corresponding index in the coding table. Since node labels/IDs can be considered a categorical variable, factorization is usually applied also on them.


\begin{figure}
{\renewcommand{\baselinestretch}{0.7}
\scriptsize
\begin{verbatim}
 *vertices  16                                           *arcs
 1 "Batagelj, Vladimir"                                  1: 1 10 1 l "authorOf"
 2 "Doreian, Patrick"                                    1: 2 10 1 l "authorOf"
 3 "Ferligoj, Anuška"                                    1: 3 10 1 l "authorOf"
 4 "Granovetter, Mark"                                   1: 1 7 1 l "authorOf"
 5 "Moustaki, Irini"                                     1: 3 7 1 l "authorOf"
 6 "Mrvar, Andrej"                                       1: 4 8 1 l "authorOf"
 7 "Clustering with relational constraint"               4: 4 14 1 l "editorOf"
 8 "The Strength of Weak Ties"                           1: 2 9 1 l "authorOf"
 9 "Partitioning signed social networks"                 1: 6 9 1 l "authorOf"
 10 "Generalized Blockmodeling"                          4: 5 11 1 l "editorOf"
 11 "Psychometrika"                                      4: 2 12 1 l "editorOf"
 12 "Social Networks"                                    3: 10 14 1 l "containedIn"
 13 "The American Journal of Sociology"                  3: 7 11 1 l "containedIn"
 14 "Structural Analysis in the Social Sciences"         3: 8 13 1 l "containedIn"
 15 "Cambridge University Press"                         3: 9 12 1 l "containedIn"
 16 "Springer"                                           2: 9 10 1 l "cites"
 *arcs :1 "authorOf"                                     2: 10 7 1 l "cites"
 *arcs :2 "cites"                                        5: 14 15 1 l "publishedBy"
 *arcs :3 "containedIn"                                  5: 11 16 1 l "publishedBy"
 *arcs :4 "editorOf"
 *arcs :5 "publishedBy"
\end{verbatim}
}
\caption{File \textbf{\texttt{bib.net}} -- bibliography network in Pajek format }
\end{figure}

\begin{figure}
   \centerline{\includegraphics[width=\textwidth]{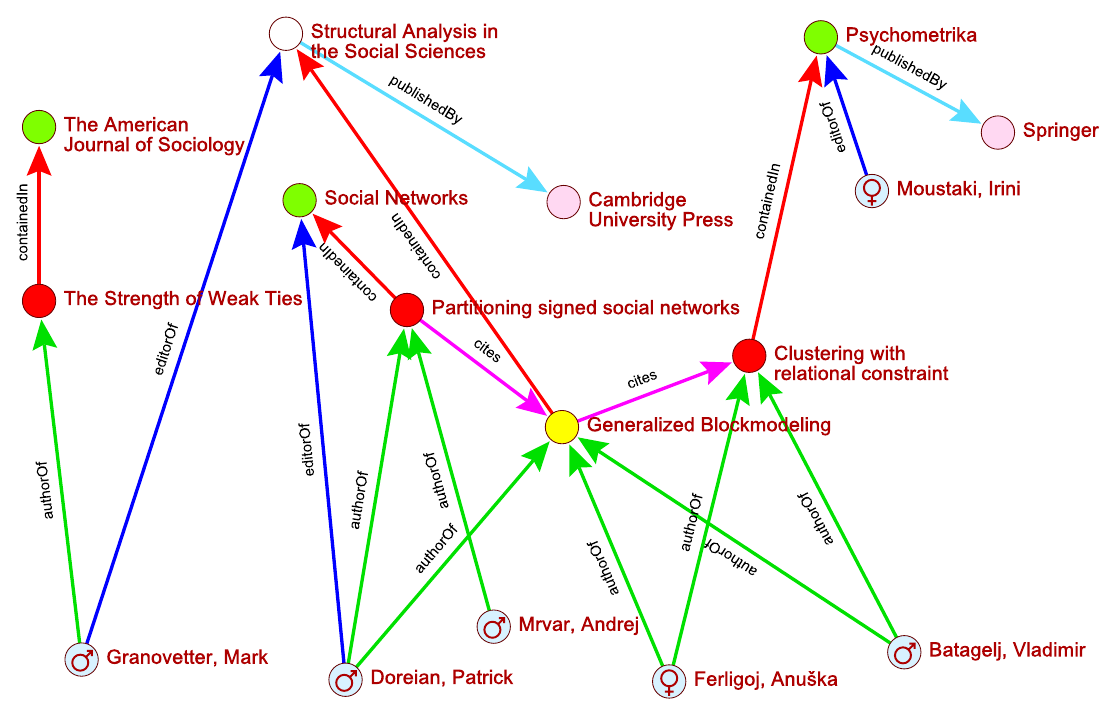}}
\caption{Bibliographic network -- picture / Pajek}
\end{figure}

This approach is used in most programs dealing with large networks. Unfortunately, the coding table is often considered as a kind of meta-data and is omitted from the description.

Be careful, in data analysis, indices start with 1, but real computer scientists start counting from 0.

Using a short program in R (see Figure 3) we transform both tables into Pajek files: a network file 
\textbf{\texttt{bib.net}}  (see Figure 4)  and partition files \textbf{\texttt{bibMode.clu}} and \textbf{\texttt{bibSex.clu}}. All the files related to the bibliographic example are available at GitHub/Bavla \cite{BavlaEx}.

\section{Nets and NetsJSON}

We were satisfied with the ''traditional'' network description until we became interested in networks with node/link properties that are not measured in standard scales (ratio, interval, ordinal, nominal), but have structured values (text, subset, interval, distribution, time series, temporal quantity, function, etc.). For describing temporal networks we initially, extending the Pajek format, defined and used the Ianus format \cite{batagelj2016algebraic}. We needed a format that could describe structured values. We could base our format on two options -- XML and JSON. We chose JSON and in 2015 started developing the NetJSON format and the Nets Python package to handle networks with structured-valued properties \cite{NetVis16, BavlaNetsJSON, BavlaNets}. 

On February 26, 2019, the format was renamed to NetsJSON because of the collision with \href{http://netjson.org/rfc.html}{http://netjson.org/rfc.html}. NetsJSON has two versions: a \keyw{basic} and a \keyw{general} version. The current implementation of the Nets library supports only the basic version. 

In addition to describing networks with structured values, NetsJSON is expected to offer the capabilities of (most) existing network description formats \cite{Bodlaj2018} (archiving, conversion) and provide input data for D3.js visualizations.

\subsection{Informal description of the basic NetsJSON format}

{\renewcommand{\baselinestretch}{0.8}\footnotesize
\begin{verbatim}
{ 
"netsJSON": "basic",
"info": {
   "org":1, "nNodes":n, "nArcs":mA, "nEdges":mE,
   "simple":TF, "directed":TF, "multirel":TF, "mode":m, 
   "network":fName, "title":title,
   "time": { "Tmin":tm, "Tmax":tM, "Tlabs": {labs} },
   "meta": [events], ...
   },
"nodes": [
   { "id":nodeId, "lab":label, "x":x, "y":y, ... },
   ***
   ]
"links": [
   { "type":arc/edge, "n1":nodeID1, "n2":nodeID2, "rel":r, ... }
   ***
   ]
}
\end{verbatim}
}

\noindent
where … are user-defined properties and *** is a sequence of such elements.

An event description can contain the following fields:
{\renewcommand{\baselinestretch}{0.8}\footnotesize
\begin{verbatim}
   {  "date": date,
      "title": short description,
      "author": name,
      "desc": long description,
      "url": URL,
      "cite": reference,
      "copy": copyright      
   }
\end{verbatim}
}
\noindent
It is intended to provide information about the "life" of the dataset -- changes, releases, uses, publications, etc.

For describing temporal networks a node element and a link element have an additional required property
\texttt{tq} -- a temporal quantity.
For an example see GitHub/Bavla/Graph/JSON/violenceU.json describing 
\href{https://raw.githubusercontent.com/bavla/Graph/refs/heads/master/JSON/violenceU.json}{Franzosi's violence network}.

The general NetsJSON format is also expected to support the description of network collections.

In recent years we also analyzed bike systems (link weight is a daily number of trips distribution), bibliographies (yearly distributions of publications or citations), and multiway networks \cite{Bikes16, batagelj2020temporal, batagelj2024cores}. It turned out that it was necessary to add another main field, \texttt{data}, to the basic NetsJSON format, in which we provide additional data about the properties of values (translations of labels in selected languages, algebraic structure \cite{Semirings}).


\section{Elements of a common network format}



Our experience with network analysis to date is summarized in the following recommendations on the elements of a common format for describing networks.

For data integrity, it makes sense to combine data and metadata into a single file. To preserve the structure of data, it makes sense to base the format on JSON, which fits well with the data structures of modern programming languages.

We would also encourage the provision, as metadata, of information about the context of the network, additional knowledge about it, articles or notebooks on its analysis, comments of users, etc. Kaggle is a good example. An improved ICON repository or Networkrepository (we disagree with their "citation request") could be the way to go. Existing metadata standards should be taken into account (\href{https://www.dublincore.org/}{Dublin Core}, \href{https://www.go-fair.org/fair-principles/}{FAIR}, \href{https://en.wikipedia.org/wiki/Schema.org}{Schema}). Data has a "life". When selecting data, its age is often important. Metadata should include at least the creation date and the last modification date.

By FAIR principles the format should support:
\begin{itemize}
\item Findability: Globally unique and persistent identifier, rich metadata.
\item Accessibility: Open, free, and universally implementable standardizes communication protocol.
\item Interoperability: Formal, accessible, shared, and proudly applicable language for knowledge representations.
\item Reusable: Metadata are richly described and associated with detailed provenance.
\end{itemize} 



The format must support all types of networks (simple, 2-mode, linked, multi-relational, multi-level, temporal). The network can contain both arcs and edges, as well as parallel links.


As mentioned earlier, using factorization produces a more concise description of the network. In cases where the node names are not too long and are readable, we sometimes want to avoid factorization. This can be achieved by using a switch that indicates whether factorization is used. We can also shorten the description length by introducing default values. If we also allow counting from 0, it makes sense to add information about the smallest index.


Long labels cause problems when printing/visualizing (parts of) networks and results. Therefore, it is useful to have abbreviated versions of labels available.

Most of the network datasets produced by network science have no node labels. Node labels are not needed if you study distributions, but they are very important in the interpretation of the obtained “important substructures”. We would encourage providing node labels, or at least some typology info in the case of privacy issues.

The common format should support descriptions of networks specific to special fields of application such as molecular graphs, genealogies \cite{white1999anthropology}, topological graph embeddings  \cite{pisanski2004representations}, etc.


We have not yet started working on a general format. It is supposed to enable descriptions of collections of networks. The question arises about the scope of validity of IDs - does the same ID in different networks represent the same or different units? This is important for operations such as union or intersection of networks. Which way to go - introducing contexts or using matchings? Maybe some ideas from the Open Archives Initiative Object Reuse and Exchange (OAI-ORE) and GraphX could be used \cite{OAIORE, gonzalez2014graphx}. An interesting option is the constructive network description -- building a network from smaller components \cite{NetML}.

Additional ideas may be found on the page ''\textit{A Python Graph API?}'' \cite{PyGraphAPI}. For now, we would leave aside descriptions of generalizations of networks (\href{https://github.com/bavla/ibm3m/tree/master/multiway}{multiway} networks and \href{https://github.com/bavla/hypernets}{hypernets}), but we must not forget about them.


\section{Conclusions}

The availability of the data used in the article enables the reproducibility and verifiability of the analyses performed. In addition, the obtained results can be verified or supplemented with other methods. When developing new methods, accessible and well-documented data are also very important - it is good to test a new method on several data sets and check whether it gives meaningful/expected results.



\newpage
\subsection{Acknowledgments}
 
The computational work reported in this paper was performed using programs R  and Pajek  \cite{de2018exploratory}. The code and data are available at Github/Bavla \cite{BavlaEx}.

V.\ Batagelj is supported in part by the Slovenian Research Agency
 (research program P1-0294, research program CogniCom (0013103)
at the University of Primorska, and research projects J5-2557, J1-2481, and J5-4596)
 and prepared within the framework of the COST action CA21163 (HiTEc).

T.\ Pisanski is supported in part by the Slovenian Research Agency 
(research program P1-0294 and research projects BI-HR/23-24-012, J1-4351, and J5-4596).

N.\ Bašić is supported in part by the Slovenian Research Agency 
(research program P1-0294 and research project J5-4596).


\bibliographystyle{splncs04}

\end{document}